\RequirePackage{fix-cm}
\documentclass[smallextended]{svjour3}       
\smartqed  
\usepackage{appendix}
\usepackage{amsmath}
\usepackage{graphicx}
\usepackage{lineno}
\usepackage{array}
\usepackage{longtable}
\usepackage{natbib}
\setcitestyle{aysep={}}

\newcommand*\patchAmsMathEnvironmentForLineno[1]{%
\expandafter\let\csname old#1\expandafter\endcsname\csname #1\endcsname
\expandafter\let\csname oldend#1\expandafter\endcsname\csname end#1\endcsname
\renewenvironment{#1}%
{\linenomath\csname old#1\endcsname}%
{\csname oldend#1\endcsname\endlinenomath}}%
\newcommand*\patchBothAmsMathEnvironmentsForLineno[1]{%
\patchAmsMathEnvironmentForLineno{#1}%
\patchAmsMathEnvironmentForLineno{#1*}}%
\AtBeginDocument{%
\patchBothAmsMathEnvironmentsForLineno{equation}%
\patchBothAmsMathEnvironmentsForLineno{align}%
\patchBothAmsMathEnvironmentsForLineno{flalign}%
\patchBothAmsMathEnvironmentsForLineno{alignat}%
\patchBothAmsMathEnvironmentsForLineno{gather}%
\patchBothAmsMathEnvironmentsForLineno{multline}%
}

\begin{document}
\title{Revisiting a drag partition model for canopy-like roughness elements}
\author{Elia Buono\and Gabriel G. Katul         \and Davide Vettori \and Davide Poggi \and Costantino Manes   
}
\institute{E. Buono \at
Dipartimento di Ingegneria dell'Ambiente, del Territorio e delle Infrastrutture, Politecnico di Torino, Torino, Italia
\email{elia.buono@polito.it}           
\and
G. Katul \at
Department of Civil and Environmental Engineering, Duke University, Durham, North Carolina, USA
\and
D. Vettori \at
Dipartimento di Ingegneria dell'Ambiente, del Territorio e delle Infrastrutture, Politecnico di Torino, Torino, Italia
\and
D. Poggi \at
Dipartimento di Ingegneria dell'Ambiente, del Territorio e delle Infrastrutture, Politecnico di Torino, Torino, Italia
\and
C. Manes\at
Dipartimento di Ingegneria dell'Ambiente, del Territorio e delle Infrastrutture, Politecnico di Torino, Torino, Italia
}

\date{Received: DD Month YEAR / Accepted: DD Month YEAR}

\maketitle

\begin{abstract}
Turbulent flows over a large surface area ($S$) covered by $n$ obstacles experience an overall drag due to the presence of the ground and the protruding obstacles into the flow. The drag partition between the roughness obstacles and the ground is analyzed using an analytical model proposed by \citet{raupach1992drag} and is hereafter referred to as R92.  The R92 is based on the premise that the wake behind an isolated roughness element can be described by a shelter area $A$ and a shelter volume $V$. The individual sizes of $A$ and $V$ without any interference from other obstacles can be determined from scaling analysis for the spread of wakes. To upscale from an individual roughness element to $n/S$ elements where wakes may interact, R92 adopted a background stress re-normalizing instead of reducing $A$ or $V$ with each element addition. This work shows that R92's approach only converges to a linear reduction in $A$ and $V$ for small $n/S$ where wakes have low probability of interacting with one another. This probabilistic nature suggests that up-scaling from individual to multiple roughness elements can be re-formulated using stochastic averaging methods proposed here. The two approaches are shown to recover R92 under plausible conditions. Comparisons between R92 and available data on blocks and vegetation-like roughness elements confirm the practical utility of R92 and its potential use in large-scale models provided the relevant parameters accommodate certain features of the roughness element type (cube versus vegetation-like) and, to a lesser extent, their configuration throughout $S$.

\end{abstract}

\section{Introduction}
The separate treatment of drag exercised by canopy-like roughness elements and the underlying surface is now needed in a plethora of science and engineering applications. These applications span
hydrodynamic behavior and sediment transport within vegetated channels \citep{huai2021flow,vargas2015effects,nepf1999drag,nepf2012hydrodynamics,aberle2013flow}, flow above urban areas \citep{grimmond1999aerodynamic,martilli2002urban,giridharan2018impact,ng2011improving,ishugah2014advances,ren2018creating,jamei2020urban,coceal2004canopy}, wind related erosion of surfaces covered by roughness elements \citep{kok2012physics,shao2008physics,raupach1993effect,ravi2010land,chappell2019minimising,okin2006multi,okin2008new}, soil-derived dust injection into the atmosphere \citep{marticorena1995modeling,zender2003mineral,shao2006review,yu2022enhanced}, deposition of aerosol particles onto vegetation and forest floor \citep{katul2010predicting,katul2011effects}, coastal dune dynamics \citep{hesp2002foredunes}, wind related tree damage \citep{gardiner2000comparison,gardiner2021wind}, to name a few.  In the early 1990s, an under-appreciated analytical approach was put forth for such drag partition over surfaces exhibiting canopy-like geometries \citep{raupach1992drag}.  This approach, labeled hereafter as R92, was derived for a deep turbulent boundary layer using scaling analysis and physical constraints on the spread of wakes behind isolated obstacles \citep{marticorena1995modeling,shao2008theory}. 
In R92, the total turbulent shear stress $\tau$ with an associated friction velocity $u_\ast$ over a rough surface of a given roughness geometry is the sum of the shear stress on all roughness elements ($\tau_R$) and the ground stress on the underlying surface ($\tau_S$). That is, the force balance per unit ground area leads to
\begin{equation}
\label{shear partition}
    \tau=\rho u_*^2=\tau_R+\tau_S,
\end{equation}
where $\rho$ is the fluid density.  In the absence of vegetation or obstacles protruding into the flow, $\tau=\rho u_*^2=\tau_S$.  The presence of roughness elements reduce the ground stress in their wake so that $\tau_S/\tau<1$. This wake effect is associated with a reduced mean velocity or, equivalently, a stress deficit behind the roughness element.  In the stress deficit representation, an effective shelter area can be defined as the area where the ground shear stress must be set to zero so as to obtain an equivalent stress deficit distributed over the entire ground area $S$. From this definition, the shelter area for a single obstacle can be defined as \citep{raupach1992drag}
\begin{equation}
\label{shelter area def}
    A=\iint\left(1-\frac{\tau_S\left(x,y\right)}{\tau_{S0}}\right)dxdy,
\end{equation}
where $\tau_S(x,y)$ is the actual ground stress at any point $(x,y)$ on $S$ and $\tau_{S0}$ is the presumed undisturbed stress, set to be equal to the actual ground stress far away from any roughness element.  Using similar arguments, an effective shelter volume can also be defined as  \citep{raupach1992drag}
\begin{equation}
\label{shelter volume def}
    V=\iiint\left(1-\frac{\phi\left(x,y,z\right)}{\phi_0}\right)dxdydz,
\end{equation}
where $\phi=\rho \alpha C_E\left(U\right)U^2$ is the local drag per unit volume on a sparse array of identical roughness elements with drag coefficient $C_E$ and frontal area per unit volume $\alpha$. As before, subscript $0$ refers to the undisturbed state or background state far away from any roughness element. In some studies, it was also interpreted as the stress value in the absence of roughness elements \citep{shao2008theory}, though the interpretation of R92 is preferred here.  

To proceed from these definitions to drag partition between the ground and the obstacles, R92 employed two plausible hypotheses. The first hypothesis is a scaling analysis that deals with the nature of the effective shelter area and shelter volume to yield
\begin{equation}
\label{shelter area}
A=c_A b h \frac{U_h}{u_\ast},
\end{equation}
and
\begin{equation}
\label{shelter volume}
V=c_V b h^2 \frac{U_h}{u_\ast},
\end{equation}
where $b$ and $h$ are the width and height of the isolated roughness element, $U_h$ is the mean flow velocity at $z/h=1$, $c_A$ and $c_V$ are unknown scaling coefficients that vary with the aspect ratio $b/h$.  In support of the first hypothesis, R92 argued that the shelter area (or volume) is dictated by how vorticity spawned by the roughness element advects and diffuses out downwind from the obstacle location. R92 argued that the vorticity produced at the obstacle advects at a velocity that scales with $U_h$ and spreads away from the longitudinal axis at a velocity that scales with $u_*$. The first hypothesis is grounded in 'text-book' scaling arguments regarding how wakes spread in boundary free flows \citep{tennekes1972first}, where $U_h$ is analogous to the so-called irrotational free stream velocity outside the wake.  The second hypothesis is that the combined effect of randomly distributed roughness elements on a surface can be obtained by random superposition of individual shelter area or volume.  The essence of this approximation is rather subtle as already pointed out in prior studies \citep{shao2008theory} and unpacking it partly motivated the present work.  In R92, the background (or far field) stress is defined as $\tau_{S0}$ in the absence of any obstacles ($i=0$).  Adding a single obstacle ($i=1$) at a random location generates a shelter area $A$ and an associated $\tau_{S1}$ that reduces $\tau_{S0}$ by a factor $(1-A/S)$. Proceeding to $i=2$ (i.e. adding another obstacle), R92 argued that the background state to use now should be $\tau_{S1}$ (already reduced from $\tau_{S0}$). This $\tau_{S1}$ is, once again, presumed to be uniformly distributed over the same ground area $S$. Adding another obstacle ($i=3$) at a random location reduces the background state, now set to $\tau_{S2}$, by another $1-(A/S)$, and so forth.  Since $A/S$ is small at each random superimposition associated with a sequential increase in $i$ or roughness element addition, the background stress state at $i-1$ can be reduced and then uniformly re-distributed over the entire area $S$.  Using these two hypotheses, the ground stress $\tau_{Sn}$ and drag force per unit area $\tau_{Rn}$ for a set of $n$ identical roughness elements randomly superimposed on $S$ are given by
\begin{equation}
\label{eq:R92_shelter_area}
\tau_{Sn}=\tau_{S0}\left(1-\frac{A}{S}\right)^n,
\end{equation}
and
\begin{equation}
\label{eq:R92_shelter_volume}
\tau_{Rn}=\frac{n\Phi}{S}\left(1-\frac{V}{Sh}\right)^n,
\end{equation}
where $S$ is presumed to be sufficiently large compared to $A$, $A$ is unaltered by successive additions of roughness elements (or increase in $i$), $\Phi=\rho C_R(U_h) bh U_h^2$ is the drag force on an isolated roughness element, with $U_h$ being the mean velocity at the canopy top, and $C_R(U_h)$ is the drag coefficient of the isolated roughness element. The ambiguity in R92 about the effect of roughness changes on the definition of $U_h$ was already raised in a prior study \citep{shao2008theory}. This ambiguity will also be considered here and shown to be less of an issue as R92 predicts $U_h/u_*$ not $U_h$.

The goal of this contribution is to answer two inter-related questions: (i) to what extent can the equations in \cite{raupach1992drag} be derived while relaxing or further clarifying some of the approximations linked to the underlying hypotheses? (ii) to what degree can the model in \cite{raupach1992drag}, or a revised version of it (as proposed here), reproduce the wealth of new data sets that include simulations and experiments over different canopy-like roughness values accrued over the last 3 decades? Thus, a practical outcome of this investigation is an 'upgrade' of model parameters that can be readily employed in conjunction with R92 for differing roughness types. 

\section{Appraisal of R92}
The appraisal begins by revising the original analysis of R92 from a discrete to a continuous case so as to accommodate the proper statistical averaging in the background stress states.  The combination of canopy stress and ground stress is then invoked to obtain the drag partition.

\subsection{Shelter Volume Analysis}
While plausible, equations \ref{eq:R92_shelter_area} and \ref{eq:R92_shelter_volume} are derived using simplified scaling of vorticity diffusion and advection. An alternative approach assumes that the turbulent kinetic energy ($TKE$) dissipation rate ($\epsilon$) scales as
\begin{equation}
\label{eq:NewScaling1}
\epsilon \sim \frac{TKE}{\tau_{d}},
\end{equation}
where $TKE$ scales as $u_\ast^2$, $u_*$ is the friction velocity and $\tau_d$ is a relaxation time describing the time it takes to dissipate the most energetic eddy. The problem then is reduced to choices for $\tau_d$.  For an isolated cylinder at moderate Reynolds numbers, $\tau_d$ may be dictated by the inverse vortex shedding frequency $f_{vs}$.  Thus, $\tau_d^{-1}$ scales with $f_{vs}=U_h St/L_{vs}$, where $St$ is the Strouhal number (assumed to be constant and equal to 0.2), $L_{vs}$ is the length scale responsible for vortex generation (e.g. width of the obstacle). However, as the roughness density increases, the main constraint on $\tau_d$ is eddy penetration into the array of roughness elements (i.e. a roughness sublayer already exists with such a roughness density).  In this case, $\tau_d$ is proportional to $l_{I}/u_*$, where $l_I$ is a characteristic scale set to either $h$ or $b$ depending on which of these two geometric scales is larger.  Analogously, $\epsilon$ can be also defined as the energy injected per unit time (and unit mass of fluid) by the work of the drag against the mean flow. Taking the velocity at the tip of the roughness element as the reference velocity leads to
\begin{equation}
\label{eq:NewScaling2}
\varepsilon\sim\frac{u_\ast^2}{\tau_d}\sim\frac{\mathrm{\Phi}U_h}{\rho V}=\frac{\rho C_R\left(U_h\right)bhU_h^2U_h}{\rho V}=\frac{C_RbhU_h^3}{V},
\end{equation}
where $V$ is the shelter volume, i.e. the volume of fluid contained in the near field of the wake generated by each roughness element 
, $\mathrm{\Phi}$ is the drag on the individual element, $C_R$ is the dimensionless drag coefficient of the roughness element with width $b$ and height $h$ (i.e. frontal area $bh$).
From equation \ref{eq:NewScaling2}, $V$ for an array of roughness elements scales as
\begin{equation}
\label{eq:NewShelterVolume}
V=C_V C_R bh\frac{U_h^3}{u_\ast^2}\tau_d,
\end{equation}
where $C_V$ is a scaling parameter. For an array of tall roughness elements, the argument in R92 can be reduced to selecting a $\tau_d$ that scales with $h/u_*$. For blunt roughness elements, $\tau_d$ may scale as $b/u_*$ or $\sqrt{b h}/u_*$. Similarly, the shelter area $A$ can be found as
\begin{equation}
\label{eq:NewShelterArea}
A= C_AV/h=C_A C_V C_R b\frac{U_h^3}{u_\ast^2} \tau_d,
\end{equation}
where $C_A$ is a scaling parameter that depends on the roughness density and shape. Since $C_R \sim{u_\ast^2}/{U_h^2}$, the R92 results are recovered when selecting a $\tau_d$ that scales as $h/u_*$.  To emphasize the role of $\tau_d$, they are repeated here as
\begin{equation}
\label{}
V=C_V b h \frac{U_h}{(u_\ast/h)},
\end{equation}
and
\begin{equation}
\label{}
A={C_AC}_Vb \frac{U_h}{(u_\ast/h)}.
\end{equation}
It is to be noted that when the roughness density is so small that no roughness sublayer can form, $\tau_d=(h/u_*)$ can be replaced with $\tau_d=b/(St U_h)$.

\subsection{Ground Stress Analysis}
The dynamically interesting features in R92 are first highlighted by contrasting them to a more intuitive starting point.  Suppose the analysis commences using a finite but very large surface area $S$ characterized by an undisturbed ground shear $\tau_{S0}$. As before, placing a single roughness element on $S$ leads to a single shelter area and a reduction to the overall ground shear stress by $\tau_{S0} A/S$ based on the definition of $A$. That is, for the $i=1$ roughness element configuration, what is outside the shelter area $A$ is assumed to be not affected by the presence of the roughness element or shelter area. This assumption means that $\tau_{S0}$ outside $A$ is the same as the background or undisturbed value whereas $\tau_S=0$ inside $A$ as before. When $\tau_S$ remains undisturbed outside the wake, a further addition of a roughness element randomly placed on $S$ will lead to an equal $\tau_{S0}A/S$ stress deficit, unless the second element is placed inside the wake area of the first.  However, this situation may be deemed unlikely (i.e. in a probabilistic manner) when $2 A/S<<1$. Extending this argument without modification to the placement of the $\it{n}$th element now leads to a ground shear stress given by
\begin{equation}
\label{eq:linear_addition}
\tau_{Sn}=\tau_{So}\left(1-\frac{nA}{S}\right). 
\end{equation}
This 'additive' approach appears plausible provided that $n A/S<<1$ and wakes have very low probability of interacting with each other because of the random placement.  This result is consistent with equation \ref{eq:R92_shelter_area} when noting that the two leading order terms in the expansion of $(1+\epsilon_o)^n$ (i.e. R92) are $1 + n \epsilon_o$ (i.e. linear additions of $A/S$) for $\epsilon_o=A/S<<1$. This difference between equation \ref{eq:linear_addition} and R92 implies that the normalization and re-distribution of the background ground shear stress state is introducing non-linearities that cannot be captured by area superposition alone.  The new physics in R92 (i.e. the background stress normalization) circumvents the naive assumption that new elements will always fall outside the existing wake areas irrespective of the number of elements already present on $S$. In fact, re-normalizing the background stress for each addition of an element accounts for the fact that  with increasing $n$, the probability of placing an element within $n A$ can no longer be ignored. To overcome this limitation the present work suggests that the proper random placement of roughness elements requires a stochastic approach to ground stress determination even though $A$ is assumed constant for each additional element.  Thus, placing a roughness element must be evaluated using its expected effect on the shear stress, which is labeled as $E(\tau_S)A$, where $E(\tau_S)$ is the expected value of $\tau_S$ when choosing a random point on $S$ already covered with $n-1$ roughness elements. For a general ground shear stress condition $\tau_S(x,y)$, $E(\tau_S)$ is given by
\begin{equation}
\label{E_tau_S}
E\left(\tau_S\right) = \lim_{k\rightarrow\infty}\left(\frac{1}{k}\sum_{i=1}^{k}\tau_S\left(x_i,y_i\right)\right).
\end{equation}
This expression is the sample mean of $\tau_{Si}$ as $k\rightarrow\infty$ points are randomly chosen over $S$. The formulation in equation \ref{E_tau_S} is equivalent to the integral mean of $\tau_S(x,y)$ over the surface $S$ because
\begin{equation}
\label{}
E\left(\tau_S\right)=\frac{1}{S}\int_{S}{\tau_SdA}.
\end{equation}
The ground stress for a series of $n$ roughness elements successively and randomly placed on surface $S$ can now be obtained. First, equation \ref{shelter area def} can be generalized by defining $\tau_{S,i}$ as the resulting ground shear stress after the $i$th element is randomly placed on $S$ and $\tau_{S,i-1}$ as the stress condition before placement of the $i$th element.  Hence, 
\begin{equation}
\label{shelter area gen}
A_i=\int_{S}\left(1-\frac{\tau_{S,i}}{\tau_{S,i-1}}\right) dA.
\end{equation}
Multiplying equation \ref{shelter area gen} by the expected value of the undisturbed ground shear state at $i-1$ results in
\begin{equation}
\label{error in t_s superimposing}
E\left(\tau_{S,i-1}\right) A_i=E\left(\tau_{S,i-1}\right)S-\int_{S}{\frac{E\left(\tau_{S,i-1}\right)}{\tau_{S,i-1}}\tau_{S,i}dA}.
\end{equation}
To recover the formulation in R92, it is necessary to assume that $E\left(\tau_{S,i-1}\right)/\tau_{S,i-1} \sim 1$ and this assumption must be satisfied for all subsequent ground shear stress fields $\tau_{S,i}$. Under this assumption,
\begin{equation}
\label{}
E\left(\tau_{S,i}\right)=E\left(\tau_{S,i-1}\right)\left(1-\frac{A_i}{S}\right).
\end{equation}
One can trace the chain of successive products all the way back to the undisturbed condition $\tau_{S0}$. For simplicity, this undisturbed state $\tau_{S0}$ must be assumed to be $E(\tau_{S0})$.  For this approximation, 
\begin{equation}
\label{eq:EB24_Sn}
\tau_{Sn}=\tau_{S0}\prod_{i=1}^{n}\left(1-\frac{A_i}{S}\right).
\end{equation}
Once again, if $A_i/S$ is a constant set to $A/S$, R92 is recovered.  The outcome in equation \ref{eq:EB24_Sn} highlights another intrinsic assumption in R92, $A_i/S$ is independent of $i$.

\subsection{Drag Analysis}
A similar approach can be used to arrive at a definition of the drag acting on the roughness elements.  That is,
\begin{equation}
\label{}
E\left(\Phi\right)=\lim_{n\rightarrow \infty}{\left(\frac{1}{n}\sum_{i=1}^{n}{\rho C_R\left(U_h\left(x_i,y_i\right)\right)bhU_h^2\left(x_i,y_i\right)}\right)}.
\end{equation}
As before, the drag is distributed over an equivalent ground area $S$.  The $U_h$ is assumed to be in equilibrium with the roughness elements placed over $S$ so that
\begin{equation}
\label{mean integral PHI}
E\left(\Phi\right)=\frac{1}{S}\int_{S}{\rho C_R\left(U_h\right) (b h) U_h^2dxdy}.	
\end{equation}
To arrive at the depth-integrated value over $h$, $\Phi$ is defined using a local drag coefficient $C_E$ that varies with the local mean velocity $U$ as  
\begin{equation}
\label{discrete PHI}
\Phi=\rho C_R bh  U_h^2=\int_{0}^{h}{\rho C_E\left(U\right)bU^2dz}.
\end{equation}
Putting together equations \ref{mean integral PHI} and \ref{discrete PHI}, a definition of $E(\Phi)$ as an integral of a discrete drag per unit volume can be derived and is given by 
\begin{equation}
\label{}
E\left(\Phi\right)=\frac{1}{S}\int_{S} \left[\frac{1}{h}\int_{0}^{h} \rho (b h) C_E\left(U\right) U^2 dz \right] dxdy.
\end{equation}
Analogous to the ground shear stress used to evaluate the drag by random superposition of $n$ roughness elements onto $S$, a general form of effective shelter volume $V_i$ of the $i$th roughness element added onto a configuration of $i-1$ roughness elements characterized by a $U_{i-1}(x,y,z)$ can be defined as
\begin{equation}
\label{shelter volume gen}
V_i=\int_{Sh}\left[1-\frac{C_E\left(U_i\right)U_i}{C_E\left(U_{i-1}\right)U_{i-1}}\right]dV.
\end{equation}
Multiplying equation \ref{shelter volume gen} by the expected value of the drag on the $i-1$ element, a relation similar to equation \ref{error in t_s superimposing} that highlights the implicit hypothesis of random super-imposition for drag contributions emerges and is given by
 \begin{equation}
 \label{}
E\left(\Phi_{i-1}\right)V_i=E\left(\Phi_{i-1}\right)Sh-\int_{Sh}{\rho bh\frac{E\left(C_E\left(U_{i-1}\right)U_{i-1}^2\right)}{C_E\left(U_{i-1}\right)U_{i-1}^2}C_E\left(U_i\right)U_i^2dV}.
\end{equation}
The hypothesis that 
\begin{equation}
\label{}
\frac{E\left[C_E(U)U^2\right]}{C_E(U)U^2}\approx 1
\end{equation}
is needed to recover the formulation for the drag on the $i$th roughness element randomly superimposed on a $i-1$  existing elements.  This formulation is given by
\begin{equation}
\label{}
E\left(\Phi_i\right)=E\left(\Phi_{i-1}\right)\left(1-\frac{V_i}{Sh}\right).
\end{equation}
It is assumed that the hydrodynamic behavior of a given random configuration of roughness elements is independent of the order in which these elements have been placed. If so, the expected drag on the $i$th element corresponds to the expected value of all other $i-1$ elements previously arranged when the $i$th element is added.  Consequently, the total drag at the $n$ element configuration is given by
\begin{equation}
\label{}
F_R=n E\left(\Phi_n\right),
\end{equation}
and
\begin{equation}
\label{eq:EB24_Vi}
F_R=n\Phi_0\prod_{i=1}^{n}\left(1-\frac{V_i}{Sh}\right).
\end{equation}
The R92 is recovered from equation \ref{eq:EB24_Vi} when $V_i/(Sh)=V/(Sh)$ is presumed constant.  In other words, as in equation \ref{eq:EB24_Sn}, an intrinsic assumption in R92 is that $V_i/(S h)$ is independent of $i$.

\subsection{Total Shear Stress}
To obtain $\tau_{Sn}$  and $\tau_{Rn}$ for n roughness elements as defined by R92, the effective shelter area and volume of each element subsequently added onto $S$ must be considered constant and equal to $A$ and $V$. This hypothesis is valid only when roughness elements are sparse and wakes of each element are similar in size to those of an isolated roughness element.  When adding sequentially a single element to $n$ elements onto $S$, the overall frontal area index $\lambda=n bh/S$ increases because $n$ increases and $S$ is constant. When a certain value of the frontal area index is achieved, R92 suggests to allow $n$ and $S$ to tend to infinity by maintaining a constant $\lambda$. This assumption leads to
\begin{equation}
\label{}
\tau_{Sn}=\tau_{S0} \left(1-\frac{\lambda A}{n bh}\right)^n,
\end{equation}
and
\begin{equation}\label{}
\tau_{Rn}=\lambda\frac{\Phi_{0}}{bh}\left(1-\frac{\lambda V}{nbh^2}\right)^n.
\end{equation}
As $n\rightarrow\infty$ and $\ S\rightarrow\infty$, this formulation tends to 
\begin{equation}
\label{}
\tau_S=\tau_{S0}\exp\left(-\frac{\lambda A}{bh}\right)
\end{equation}
and
\begin{equation}\label{}
\tau_R=\lambda\frac{\Phi_{0}}{bh}\exp\left(-\frac{\lambda A}{bh}\right).
\end{equation}
Here, the identity 
\begin{equation}
\label{}
\lim_{n \to +\infty} \left(1 + \frac{x}{n} \right)^n = \exp\left(x\right)
\end{equation}
was employed.  Recalling that drag acting on the isolated element is $\Phi_0=\rho C_R (bh) U_h^2$ and the undisturbed ground shear stress can be generically expressed as $\tau_{So}=\rho C_S U_h^2$ and putting all these together yields
\begin{equation}
\label{all together 1}
\tau=\rho C_S U_h^2 \exp \left(-\frac{\lambda A }{bh} \right) +\rho \lambda C_R U_h^2 \exp \left( -\frac{\lambda V }{bh^2} \right).
\end{equation}
Upon setting $V=Ah$ and $A_f=bh$, equation \ref{all together 1} simplifies to
\begin{equation}
\label{final total shear}
\frac{1}{\gamma^2}=(C_S + \lambda C_R) \exp \left(-\lambda\frac{ A }{A_f} \right),
\end{equation}
where $\gamma=U_h/u_*$. Equation \ref{final total shear} remains valid irrespective of the precise definition of shelter area. From equation \ref{shelter area}, $A/A_f=c_A(U_h/u_*)$ is constant and equation \ref{final total shear} can be formulated as 
\begin{equation}
\label{final form}
\frac{1}{\gamma^2}=(C_S + \lambda C_R) \exp \left(-c_A\lambda \gamma\right).
\end{equation}
Equation \ref{final form} can be expressed as
\begin{equation}
\label{final form2}
Y \exp (-Y) = B_o; Y= \frac{c \lambda \gamma}{2}; B_o=\frac{1}{\sqrt{C_s + \lambda C_R}} \frac{c_A \lambda \gamma}{2}.
\end{equation}
The solution to this algebraic equation (i.e. $Y=f(B_o)$) can be expressed in terms of the Lamberts $W$ function.  However, the Lamberts $W$ function (or its related function - the product logarithm) cannot be expressed in terms of elementary functions and numerical methods must be employed to solve this equation. Few properties of this equation have already been highlighted by R92 and are briefly repeated: no solutions for $B_o>\exp(-1)$ exist, a single solution for $B_o=\exp(-1)$ (i.e. $Y=1$) exist, and two solutions for $B_o<\exp(-1)$ also exist.  The two solutions are $Y_1<1$ and $Y_2>1$.  The physically realistic solution is the $Y_1<1$.  

The drag partition problem is now reduced to inferring $C_S$ and $C_R$ (as well as $c_A$) from measured $\gamma$ and $\lambda$.  This inference is conducted using nonlinear optimization - matching equation \ref{final form} to measured $\gamma$ and $\lambda$ while setting $C_S$, $C_R$, and $c_A$ to constants whose values are optimized from least-squares analysis. 

\subsection{A Simplified Expression for Drag Partition}
Before proceeding to the experimental evaluation, a number of comments can be made about a popular variant of R92.  From definitions and provided that $A/S = V/(S h)$ remains constant, a simpler version of R92 discussed in other studies \citep{raupach1993effect,shao2008theory} can be directly recovered. Equations \ref{eq:R92_shelter_area} and \ref{eq:R92_shelter_volume} can be expressed as 
\begin{equation}
\label{eq:iso}
\tau_{Sn}=\rho C_S U_h^2\left(1-\frac{A}{S}\right)^n;~ \tau_{Rn}=\rho C_R \lambda U_h^2\left(1-\frac{A}{S}\right)^n=(\beta \lambda) \tau_{Sn},
\end{equation}
where $\beta=C_R/C_S$. It directly follows that the drag partition problem of $\tau$ into its two components for $n$ elements reduces to
\begin{equation}
\label{eq:iso2}
\frac{\tau_{Sn}}{\tau}=\frac{1}{1+\beta \lambda}; ~ \frac{\tau_{Rn}}{\tau}=\frac{(\beta \lambda)}{1+\beta \lambda}.
\end{equation}
This equation has the desirable features that for large $\beta \lambda$, $(\tau_{Sn}/\tau) \rightarrow 0$ and $\tau_{Rn}/\tau \rightarrow 1$ consistent with logical expectations that the total drag is carried by drag on the $n$ elements and not by the ground.  Conversely, when $\beta \lambda \rightarrow 0$,  $(\tau_{Sn}/\tau) \rightarrow 1$ and $\tau_{Rn}/\tau \rightarrow 0$, again as expected.  The expressions in equation \ref{eq:iso2} can be stated as
\begin{equation}
\label{eq:iso3}
\gamma^2=\frac{1}{1+\beta \lambda} \frac{1}{C_S} \left(1-\frac{A}{S}\right)^{-n}.
\end{equation}
For $n A/S$ being small, 
\begin{equation}
\label{eq:iso4}
\frac{1}{\gamma^2}=(1+\beta \lambda){C_S} \left(1-n\frac{A}{S}\right)=(C_S+C_R \lambda) \left(1-c_A \lambda \gamma \right).
\end{equation}
That is, equation \ref{final form} becomes reasonably approximated by its linear form in equation \ref{eq:iso4} for small $n A/S$.  This simplified version illustrates the quadratic dependence of $\gamma^{-2}$ on $\lambda$. 

\section{Experimental Evaluation}
Data presented in R92 along with recent experiments are here combined to evaluate R92 across different roughness types (see details in Table~\ref{Tab:datasets}). The data sets refer to the most studied obstructions: prismatic objects referred to as “cubes” \citep{walter2012shear,macdonald1998improved,yang2016exponential,macdonald2000modelling} and plant-shaped objects referred to as "plants", both artificial \citep{raupach1992drag,walter2012shear,poggi2004effect,kang2019experimental} and real \citep{wolfe1996shear,lancaster1998influence}, and from laboratory and field experiments as well as numerical simulations \citep{yang2016exponential}. The data sets cover the following ranges of $\lambda$: from $3\times10^{-4}$ to $5$ for real plants, from $3.7\times10^{-3}$ to $0.2$ for artificial plants and from $3.9\times10^{-3}$ to $9.1\times10^{-2}$ for cubes. 

\begin{table}[!ht]
    \caption{\label{Tab:datasets}Summary of collected experimental and simulation results from the literature (* field experiment, ** numerical simulation). Here, the Data ID is used as legend in Figures  \ref{Fig:raupach_plot}, \ref{Fig:linear_comp}, \ref{Fig:mean parameters plants}, \ref{Fig:mean parameters cubes}, \ref{Fig:odd cubes}.  The $h$ is the roughness element height, $b$ is the roughness element average breath $b=A_f/h$, $A_f$ is the roughness element frontal area, $\lambda$ is the frontal area index, $\gamma$ is the ratio $U_h/u_*$ and $C_S$ is the ground surface drag coefficient ($'$ is used when not provided by the authors).}
    \centering
    \begin{tabular}{|c|c|c|c|c|c|c|c|c|c|}
    \hline
        Reference & Data ID & Type & $h\  [mm]$ & $b\  [mm]$ & $\lambda \times 10^3$ & $\gamma$ & $C_S\times 10^3$  \\ \hline
        \cite{raupach1992drag} & RTE & Plant & 6 & 6 & 10.8-182 & 12.5-5.4 &  3.0  \\
            & OL & Cube & 4.7 & 4.8 & 3.9-15.6 & 15.4-11.2 &  3.0  \\
            & GJR* & Plant & - & - & 253-5000 & 3.9-2.6 &  3.0  \\ \hline
        \cite{walter2012shear} & WA\_pla & Plant & 100 & 40 & 15-202 & 14.6-6.6 & 1.8  \\ 
            & WA\_cub & Cube & 80 & 40 & 16.7-173 & 11.8-5.0 & 1.9  \\ \hline
        \cite{macdonald2000modelling} & MCD\_al & Cube & 100 & 100 & 50-330 & 6.9-3.9 & 2.0'  \\ 
            & MCD\_st & Cube & 100 & 100 & 50-330 & 5.6-2.8 & 2.0'  \\ \hline
        \cite{wolfe1996shear}* & WLF & Plant & 820-1200 & 1100-1320 & 0.3-232 & 13.5-6.2 & 5.2  \\ \hline
        \cite{lancaster1998influence}* & LAN & Plant & 90-110 & 90-110 & 39-212 & 11.8-5.8 & 4.0'  \\ \hline
        \cite{yang2016exponential}** & YAN\_al & Cube & 25 & 25 & 30-250 & 8.2-3.4 & 2.0' \\ 
           & YAN\_st & Cube & 25 & 25 & 30-250 & 8.3-3.1 & 2.0' \\ \hline
        \cite{macdonald1998improved} & MCDb\_al & Cube & 100 & 100 & 48-910 & 7.8-4.0 & 2.0' \\ 
            & MCDb\_st & Cube & 100 & 100 & 48-910 & 6.3-3.6 & 2.0' \\ \hline
       \cite{poggi2004effect} & PGG\_04 & Plant & 120 & 4 & 32.2-515 & 17.2-3.7 & 2.0' \\ \hline
        \cite{kang2019experimental} & KAN\_19 & Plant & 3.5-15.0 & 5.2-13.2 & 3.7-156 & 17.0-4.6 & 2.4 \\ \hline
        \cite{li2022bridging}** & LI\_22 & Cube & - & - & 80-438 & 6.7-1.5 & 2.0' \\ \hline
        \cite{placidi2015effects} & PLA\_15 & Cube & 7.8-15.6 & 11.4 & 120-240 & 5.5-1.4 & 2.0' \\ \hline

    \end{tabular}
\end{table}

\begin{figure}
\centering
\noindent\includegraphics[width=20pc]{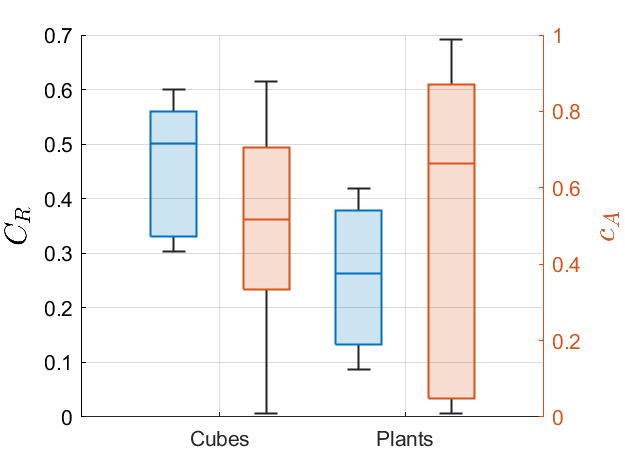}
\caption{Box plot of the optimized drag coefficient $C_R$ (left ordinate, blue boxes) and scaling coefficient $c_A$ (right ordinate, red boxes) obtained from the non linear regression for the two roughness shapes (cubes and plants) and all data sets.}
\label{Fig:C_R boxplot}     
\end{figure}

\begin{figure}
\centering
\noindent\includegraphics[width=17.4cm]{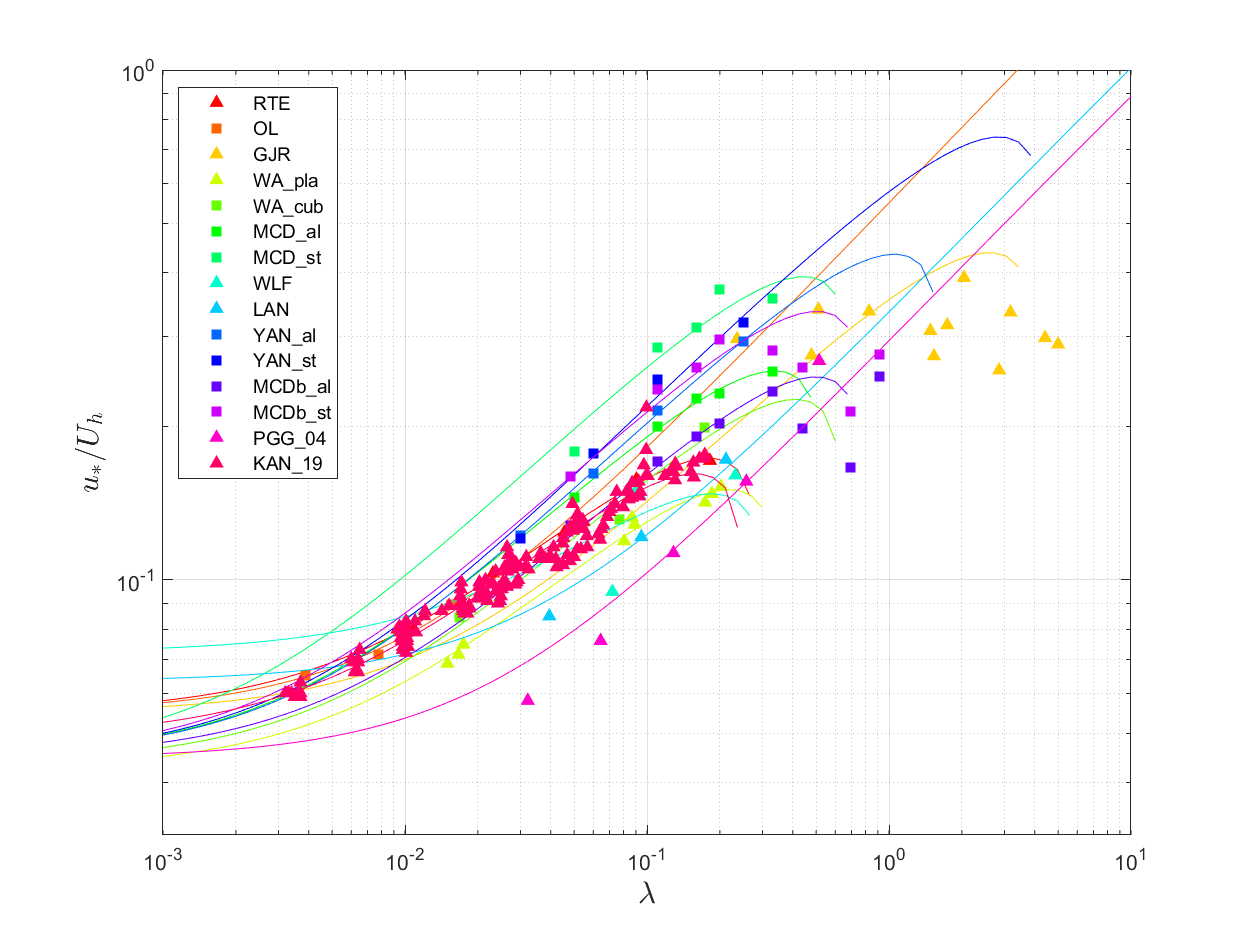}
\caption{The variation of $u_*/U_h$ (ordinate) with frontal area index $\lambda$ (abscissa) for plants (triangles) and cubes (squares). The R92 (i.e. equation \ref{final form}) is shown in solid lines for the optimized parameters.
}
\label{Fig:raupach_plot}     
\end{figure}

\begin{figure}
\centering
\noindent\includegraphics[width=17.4cm]{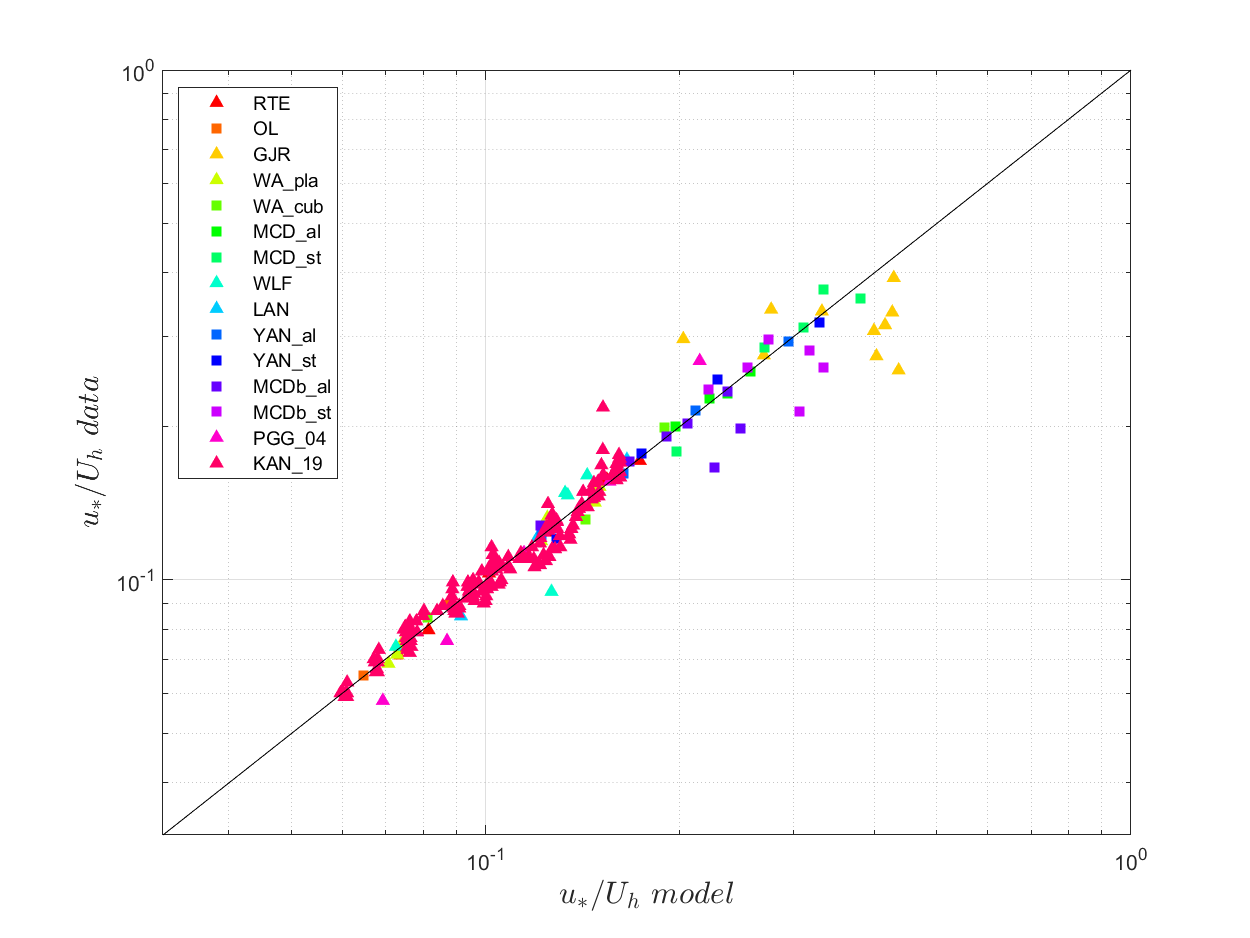}
\caption{Evaluation of R92 for predicting $u_*/U_h$.  The experimental and numerical measurements of $u_*/U_h$ (ordinate) are compared with predictions of $u_*/U_h$ (abscissa) using the optimized parameters for plants (triangles) and cubes (squares).}
\label{Fig:linear_comp}     
\end{figure}

In the majority of the experiments $\lambda$ and $\gamma$ were measured and combined with equation \ref{final form} to obtain the unknowns $C_R$ and $c_A$ by means of non-linear least squares regression. The third degree of freedom (i.e. $C_S$) in equation \ref{final form} was fixed at values provided in each study to ensure robust regression fits. When a value of $C_S$ was not provided, we arbitrarily set $C_S=2\times 10^{-3}$ for laboratory experiments and $C_S=4\times 10^{-3}$ for field experiments.  The nonlinear regression analysis was conducted using MATLAB's (Mathworks, Natick, Massachusetts, USA) built in function "lsqnonlin". The optimized values of $C_R$ and $c_A$ are summarized in Table~\ref{Tab:results} and Figure \ref{Fig:C_R boxplot}. The optimized curves are presented in Figure \ref{Fig:raupach_plot} and a comparison between measured and modeled $u_*/U_h$ is shown in Figure \ref{Fig:linear_comp}. 

\begin{table}[!ht]
    \centering
        \caption{\label{Tab:results}Summary of published experimental and simulation results from literature (* field experiment, ** numerical simulation). Here the Data ID is the legend code used in Figures \ref{Fig:raupach_plot}, \ref{Fig:linear_comp}, \ref{Fig:mean parameters plants}, \ref{Fig:mean parameters cubes}, \ref{Fig:odd cubes}.  As before, the $C_S$ is the fixed ground surface drag coefficient ($'$ is used when not provided by the authors), $C_R$ and $c_A$ are the free parameters obtained from nonlinear regression and $R^2$ is the coefficient of determination.}
    \centering

    \begin{tabular}{|c|c|c|c|c|c|}
    \hline
        Data ID & Rough Type & $C_S$ & $C_R$ & $c_A$ & $R^2$ \\ \hline
        RTE & plant & 0.003 & 0.42 & 0.92 & 1.00 \\ \hline
        OL & cube & 0.003 & 0.30 & 0.01 & 0.98 \\ \hline
        GJR* & plant & 0.003 & 0.20 & 0.17 & -4.91 \\ \hline
        WA\_pla & plant & 0.0018 & 0.26 & 0.66 & 0.99 \\ \hline
        WA\_cub & cube & 0.0019 & 0.33 & 0.54 & 0.97 \\ \hline
        MCD\_al & cube & 0.002' & 0.52 & 0.75 & 0.99 \\ \hline
        MCD\_st & cube & 0.002' & 0.93 & 0.88 & 0.92 \\ \hline
        WLF* & plant & 0.0052 & 0.26 & 0.72 & 0.75 \\ \hline
        LAN* & plant & 0.004 & 0.11 & 0.01 & 0.97 \\ \hline
        YAN\_al** & cube & 0.002' & 0.48 & 0.41 & 1.00 \\ \hline
        YAN\_st** & cube & 0.002' & 0.52 & 0.26 & 0.98 \\ \hline
        MCDb\_al & cube & 0.002' & 0.32 & 0.49 & 0.32 \\ \hline
        MCDb\_st & cube & 0.002' & 0.60 & 0.66 & 0.19 \\ \hline
        PGG\_04 & plant & 0.002' & 0.09 & 0.01 & 0.93 \\ \hline
        KAN\_19 & plant & 0.0024 & 0.42 & 0.99 & 0.96 \\ \hline
        LI\_22 & cube & 0.002' & 1.22 & 1.17 & 0.38 \\ \hline
        PLA\_15 & cube & 0.002' & 1.57 & 1.56 & 0.091 \\ \hline
    \end{tabular}

\end{table}


The effectiveness of R92 is evaluated using the coefficient of determination $R^2$, which is reported in Table \ref{Tab:results} for all data sets employed.  Equation \ref{final form} provides a good ($R^2>0.9$) description of $u_*/U_h$ across most data sets. Consistent with model assumptions, equation \ref{final form} appears less effective for high values of $\lambda$. Lower $R^2$ values were also obtained for field experiments, which are typically characterized by large uncertainties and heterogeneous morphology (see GJR, WLF, MCDb\_al and MCDb\_st in Table \ref{Tab:results}). 

\begin{figure}
\centering
\noindent\includegraphics[width=17.4cm]{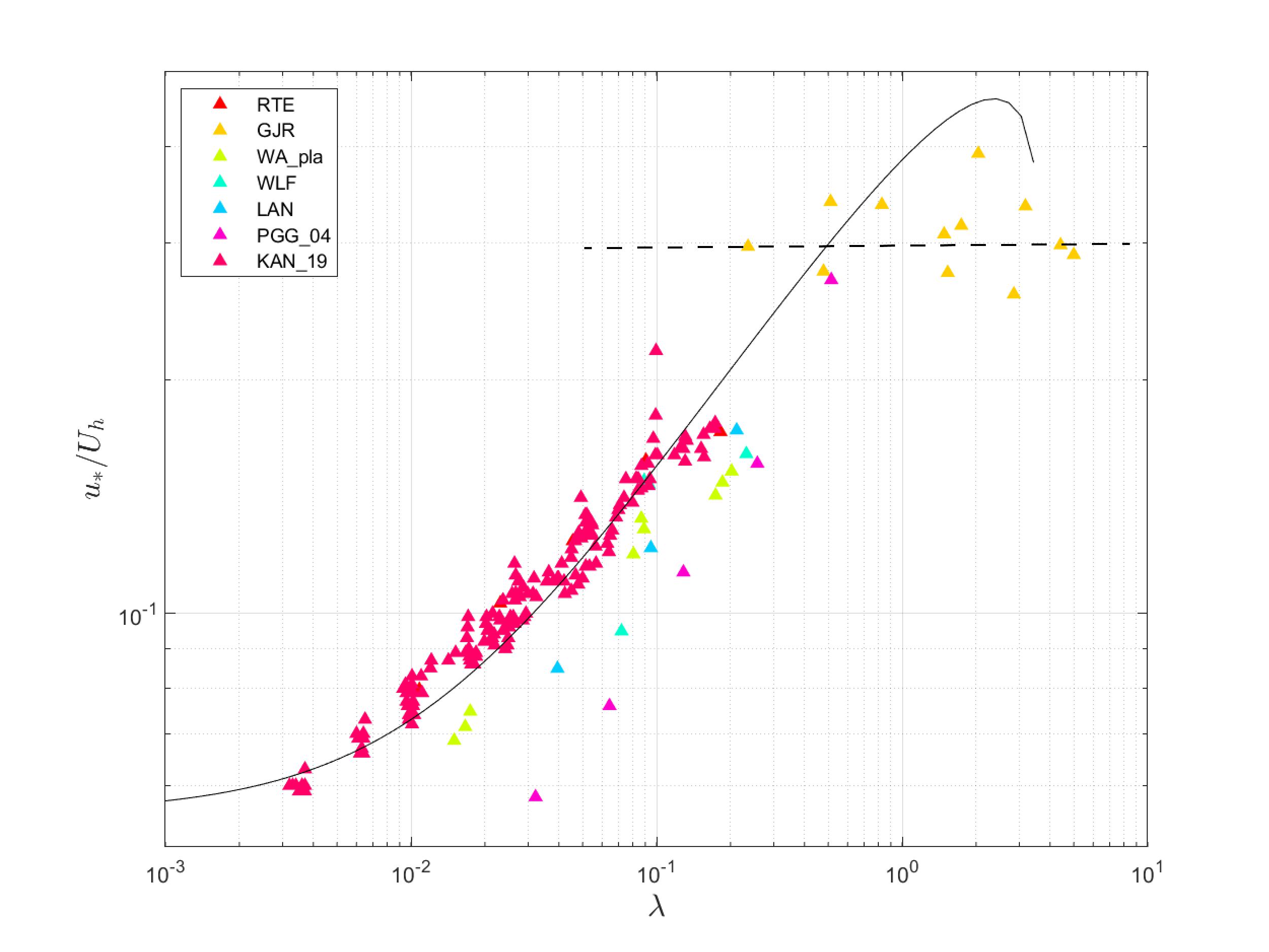}
\caption{The experimental results of $u_*/U_h$ (ordinate) along with the frontal area index $\lambda$ (abscissa) for plants compared with the mean parameterized equation \ref{final form} ($C_S=0.002$, $C_R=0.24$,$c_A=0.19$)}
\label{Fig:mean parameters plants}     
\end{figure}

\begin{figure}
\centering
\noindent\includegraphics[width=17.4cm]{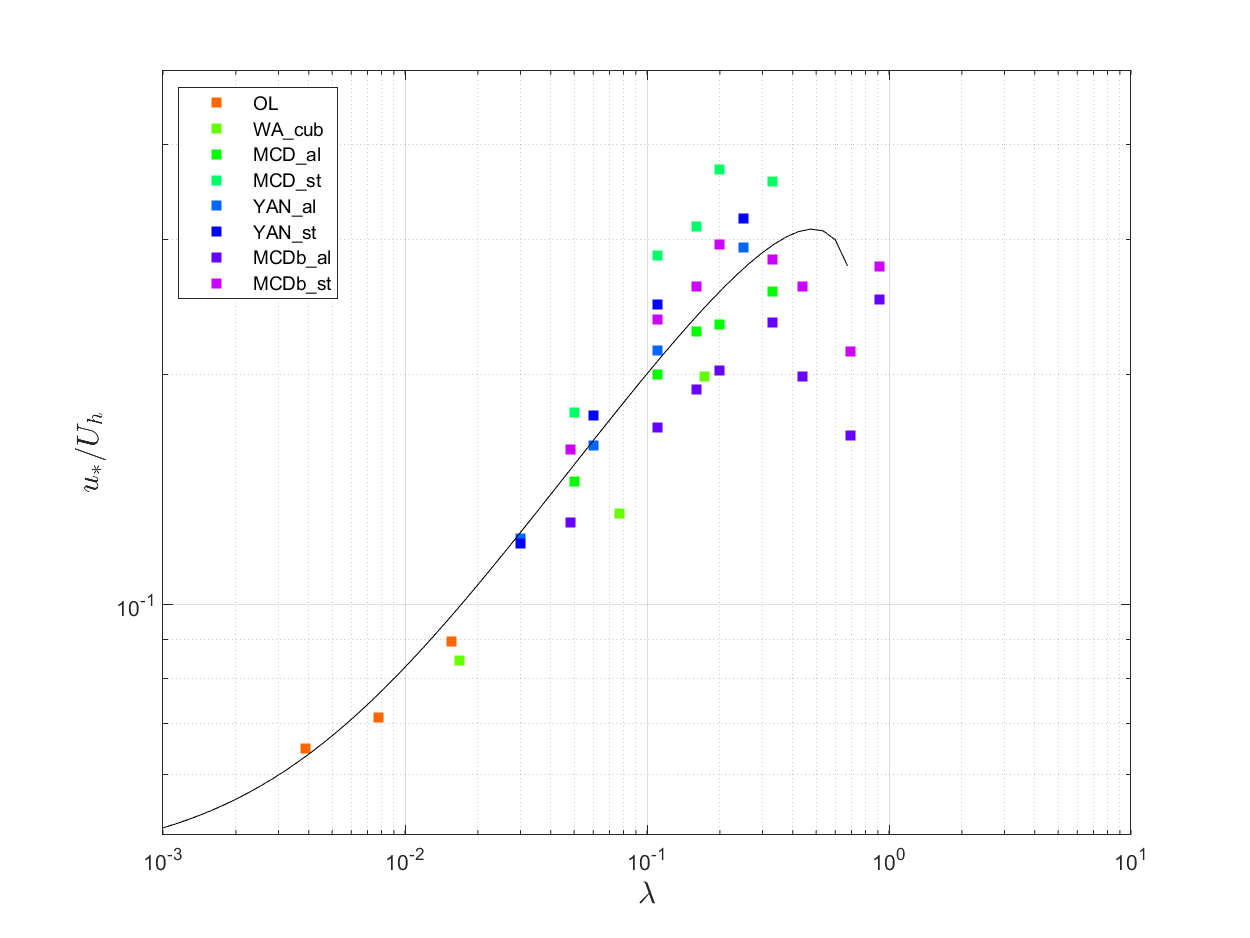}
\caption{The experimental and numerical results of $u_*/U_h$ (ordinate) along with the frontal area index $\lambda$ (abscissa) for cubes compared with the mean parameterized equation \ref{final form} ($C_S=0.002$, $C_R=0.53$,$c_A=0.63$)}
\label{Fig:mean parameters cubes}     
\end{figure}

As shown in Figure \ref{Fig:C_R boxplot}, the variability in $c_A$ exceeds that of $C_R$, with no significant differences observed between cubes and plants. The variability in $C_R$ across experiments could be foreshadowed given that the dependence of $C_R$ on the Reynolds number was not considered. Despite a higher degree of heterogeneity in shapes, the range of $C_R$ for plants is surprisingly similar to that of cubes. However, the parameter $c_A$ exhibits a wider spread suggesting that the definitions of $A$ and $V$ in R92 might oversimplify quantities that are intricately influenced by numerous factors beyond those already incorporated in R92 (i.e. $\lambda$ and $\gamma$), such as planar area index, aspect ratio, frontal solidity, internal porosity of rough elements, the roughness of the surface skin, or the dissipation rate of turbulent kinematic energy to name a few.
To move R92 to an operational form, the data were also compared to predictions from equation \ref{final form} using values of $C_R=$ 0.24 and $c_A=$ 0.19 for plants and $C_R=$ 0.53 and $c_A=$ 0.63 for cubes, obtained by nonlinear regression off all data from each roughness element type, regardless of the data set. This comparison is shown in Figure \ref{Fig:mean parameters plants} for plants and Figure \ref{Fig:mean parameters cubes} for cubes. Remarkably, despite being computed with approximate parameters, the R92 approach closely aligns with the experimental results. The $R^2$ for the data modeled with those $C_R$ and $c_A$ was 0.79 for cubes and 0.86 for vegetation, offering a pragmatic approach to large-scale models. This parameterization establishes an upper bound for the R92 model, which appears to diminish in effectiveness for $\lambda \geq$ 0.3, observed for both plants and cubes and in agreement with previous works \citep{katul2004one}.

\begin{figure}
\centering
\noindent\includegraphics[width=20pc]{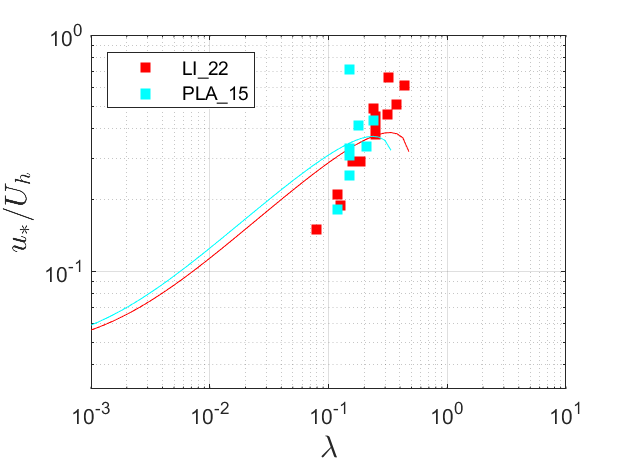}
\caption{Relation between $u_*/U_h$ and $\lambda $ for cuboid experiments and LES runs where the frontal area and the planar area both vary.}
\label{Fig:odd cubes}     
\end{figure}

A comparison between R92 and two other studies for cuboids is shown separately in Figure \ref{Fig:odd cubes}.  These studies, which cover LES \citep{li2022bridging} and laboratory experiments with LEGO as obstacles \citep{placidi2015effects} are separated from the other cube studies because they explore how $u_*/U_h$ vary when the planar area density is altered for a preset $\lambda$.  Also, $\lambda>0.1$ in both studies, thus approaching the range where R92 acknowledged potential failure ($\lambda=0.3$) due to wake interference with obstacle placement. While these two studies reasonably agree with each other, R92 fails to predict $u_*/U_h$ from $\lambda$ alone.  This failure underscores the role of the planar area index (i.e. length and width are the defining obstacle geometry) in altering the shelter area beyond width and height (considered in the roughness area), which is neglected in R92.  Nonetheless, the collapse of the two data sets with $\lambda$ is simultaneously surprising and encouraging, and perhaps hints that R92 may be modified to include such effects.  This modification is better left for a future study when more data sets will be available.      

\section{Conclusions}
A drag partition model (R92) that separates the overall stress into  contributions from roughness obstacles and from the ground was analyzed and evaluated with data collected over the past 3 decades.  Revisions proposed here do not alter the main findings of R92, but provide a stochastic framework for analyzing the effects of random obstacle additions.  Under some restrictive conditions, these revisions recover the re-normalized background stress proposed by R92 but in an ensemble-averaged sense.  Such a stochastic averaging formalizes the up-scaling from individual to multiple roughness elements in R92 and clarifies some of the assumptions behind the stress re-normalization. An important condition is that the roughness elements are distributed homogeneously throughout the surface area and do not form clusters/patches, in which case wakes would interact. Agreement between R92 and published data on cubes and vegetation-like roughness elements is good for most data sets (many of which were collected after R92 was published).  A poor performance was recorded for high roughness densities where wake interference between shelter areas is to be expected (both in R92 and its stochastic version proposed here).  The mean value of $C_R=0.24$ for plants is close to the value reported by R92 ($C_R=0.3$).  Using a single $C_R$ and $c_A$ for plants ($C_R$=0.24 and $c_A$=0.19) and cubes ($C_R$=0.53 and $c_A$=0.63) captures the $u_*/U_h$ variations with frontal area density $\lambda$.  For $\lambda \geq$ 0.3, the modeled $u_*/U_h$ degrades when using R92 without modification -- though selecting a constant value of $u_*/U_h$=0.3 appears superior as already proposed by R92 and shown to be reasonable for many terrestrial vegetation studies \citep{katul2004one}.  For high $\lambda$ and for cases where the planar area density variations are large at a preset $\lambda$, the R92 performance leaves much to be desired and is worth exploring further. Despite this limitation, it may be surmised that the simplicity, the theoretical foundations, and the broad experimental support for R92 makes it practical to implement in large-scale weather forecasting and climate related models.  

\bibliographystyle{spbasic_updated}   
\bibliography{sample_library} 
\newpage
\end{document}